\documentclass[twocolumn,prc,superscriptaddress,unsortedaddress,showpacs,preprintnumbers,amsmath,amssymb]{revtex4}
\usepackage{tipa}

\usepackage{graphicx}
\usepackage{dcolumn}
\usepackage{bm}

\def\beq{\begin{equation}}
\def\eeq{\end{equation}}
\def\bea{\begin{eqnarray}}
\def\eea{\end{eqnarray}}

\def\fun#1#2{\lower3.6pt\vbox{\baselineskip0pt\lineskip.9pt
  \ialign{$\mathsurround=0pt#1\hfil##\hfil$\crcr#2\crcr\sim\crcr}}}

\begin{document}

\preprint{}

\title{Effects of tensor interaction on pseudospin energy splitting and shell correction}

\author{J. M. Dong}
\affiliation{Research Center for Hadron and CSR Physics, Lanzhou University and Institute of Modern Physics of CAS, Lanzhou, China}
 \affiliation{Institute
of Modern Physics, Chinese Academy of Sciences, Lanzhou 730000,
China} \affiliation{Graduate University of Chinese Academy of
Sciences, Beijing 100049, China} \affiliation{China Institute of
Atomic Energy, P. O. Box 275(10), Beijing 102413, China}
\affiliation{School of Nuclear Science and Technology, Lanzhou
University, Lanzhou 730000, China}
\author{W. Zuo}\email[ ]{zuowei@impcas.ac.cn}
\affiliation{Research Center for Hadron and CSR Physics, Lanzhou University and Institute of Modern Physics of CAS, Lanzhou, China}
\affiliation{Institute of Modern Physics, Chinese Academy of
Sciences, Lanzhou 730000, China}
 \affiliation{School of Nuclear
Science and Technology, Lanzhou University, Lanzhou 730000, China}
\affiliation{Center of Theoretical Nuclear Physics, National
Laboratory of Heavy Ion Accelerator of Lanzhou, Lanzhou 730000}
\author{J. Z. Gu}\email[ ]{gujianzhong2000@yahoo.com.cn}
\affiliation{China Institute of Atomic Energy, P. O. Box 275(10),
Beijing 102413, China} \affiliation{Center of Theoretical Nuclear
Physics, National Laboratory of Heavy Ion Accelerator of Lanzhou,
Lanzhou 730000}
\author{Y. Z. Wang}
\affiliation{China Institute of Atomic Energy, P. O. Box 275(10),
Beijing 102413, China}
\author{L. G. Cao}
\affiliation{Institute of Modern Physics, Chinese Academy of
Sciences, Lanzhou 730000, China}
\author{X. Z. Zhang}
\affiliation{China Institute of Atomic Energy, P. O. Box 275(10),
Beijing 102413, China}

\begin{abstract}
In the framework of a Skyrme-Hartree-Fock approach combined with BCS
method, the role of the tensor force on the pseudospin energy
splitting for tin isotope chain is investigated. The tensor force
turns out to obviously affect the pseudospin energy splitting of the
spin-unsaturated nuclei. Since the tensor force shifts the
single-particle levels, it modifies the single-particle level
density and the shell correction energy thereof. The influence of
the tensor interaction on shell correction energy is considerable
according to our analysis taking a magic nucleus $^{132}$Sn as well
as a superheavy nucleus $^{298}114$ as examples. This modification
of the shell correction energy due to the tensor component affects
the stability of the superheavy nuclei.

\end{abstract}

\maketitle
\section{Introduction}\label{intro}\noindent
Properties of exotic nuclei are not only interesting from the
viewpoint of nuclear structure, but also have important implications
in nuclear astrophysics, such as in nova and supernova explosions,
X-ray bursts associated with explosive hydrogen burning, rapid
proton capture processes, etc. In view of the recent rapid progress
in the discovery of exotic nuclei driven by the advent of the new
generation of radioactive ion beam facilities, an important problem
is to understand how the shell structure evolves from stable to
exotic nuclei, in which the role of the tensor force has received
wide attention. The effect of the tensor force on the shell
evolution has been explored by Otsuka and his collaborators within
the shell model. They found that the tensor force strongly affects
the evolution of the single-particle energy level spacing and shell
structure \cite{Otsuka1}.

In the case of Skyrme interactions, a zero-range tensor component
was actually present in the original papers of Skyrme \cite{SKY1}.
However, this component was omitted when realistic Skyrme parameter
sets were determined, and it has been systematically neglected for
several decades in practical applications. Until very recently, the
tensor component was added to the existing Skyrme parametrizations
or included by fitting new parametrizations to investigate the
contributions of the tensor force to the spin-orbit splitting
\cite{SO1,SO2,SO3}, evolution of single-particle energies
\cite{SO3,single1,single2,single3}, shell evolution
\cite{SO3,shell1,shell2}, stability of superheavy nuclei \cite{SHN},
binging energy \cite{BE1,BE2}, nuclear deformability \cite{BE2,BE3},
properties of excited states \cite{FF1,FF2}, and the spin and
spin-isospin instabilities of nuclear matter \cite{cao}.

The concept of pseudospin was supported by experimental observations
about 40 years ago that the single-nucleon doublet levels with
quantum numbers ($n_{r}$, $l$, $j=l+1/2$) and ($n_{r}-1$, $l+2$,
$j=l+3/2$) lie very close in energy, where $n_{r}$, $l$, $j$ are the
single nucleon radial, orbital, and total angular momentum quantum
numbers, respectively. This single-nucleon doublet pair can be
relabelled as a pseudospin doublet: ($\widetilde{n}_{r}=n_{r}-1$,
$\widetilde{l}=l+1$, $\widetilde{j}=\widetilde{l}\pm 1/2$). Then the
two states in the doublet are almost degenerate with respect to
pseudospin. This symmetry has been used to explain a number of
phenomena in nuclear structure including the deformation \cite{DD},
superdeformation \cite{SD}, magnetic moment \cite{MR1,MR2} and
identical bands \cite{band1,band2,band3}. It is also pointed out
that the conservation of the pseudospin symmetry plays an essential
role in stabilizing the neutron halo structure and generating the
neutron shell effects when the neutron number approaches the neutron
drip line \cite{LWH}. The details about the pseudospin symmetry can
be found in Refs. \cite{PS1,PS2,PS3}. It is interesting to discuss
the influence of the tensor force on the pseudospin symmetry. In
this work, we shall discuss the effect of the tensor component of
the two-body effective interaction on the pseudospin energy
splitting by employing a Skyrme-Hartree-Fock approach, taking a long
chain of Sn isotopes as an example. In addition, the effect of
tensor interaction on shell correction energy will be investigated.
The liquid drop model explains well the gross features of nuclear
fission and mass, however, there are systematic deviations from the
smooth mass. The shell correction, which is a fluctuation in the
binding energy, can be supplemented to the liquid-drop model to
improve the description of the nuclear masses and fission barriers
\cite{P1,HF,P2,P3,P4,P5,P6}. For superheavy nuclei, as an important
property, the shell correction energy is widely investigated because
it is relevant to the stability of a superheavy nucleus. The
strutinsky method \cite{ST}, which is widely used today, will be
employed to extract the shell correction energy here. With the
single-particle spectra obtained from the Skyrme-Hartree-Fock
approach, the shell correction energies for heavy and superheavy
nuclei can be calculated and one can analyze the effect of the
tensor force.

This paper is organized as follows: In section II, a brief
theoretical method is presented. The effects of the tensor
interaction on the pseudospin energy splitting and shell correction
are discussed in sections III and IV, respectively. Finally, a brief
summary is provided in section V.

\section{Tensor force in the framework of the Skyrme interaction}\label{intro}\noindent
The tensor part of the Skyrme effective interaction is written as
\cite{SKY1}
\begin{eqnarray}
v_{T} &=&\frac{T}{2}\left[ (\mathbf{\sigma }_{1}\cdot \mathbf{k}^{\prime })(%
\mathbf{\sigma }_{2}\cdot \mathbf{k}^{\prime })-\frac{1}{3}\mathbf{k}%
^{\prime 2}(\mathbf{\sigma }_{1}\cdot \mathbf{\sigma }_{2})\right] \delta (%
\mathbf{r}_{1}-\mathbf{r}_{2})  \notag \\
&&+\frac{T}{2}\left[ (\mathbf{\sigma }_{1}\cdot \mathbf{k})(\mathbf{\sigma }%
_{2}\cdot \mathbf{k})-\frac{1}{3}(\mathbf{\sigma }_{1}\cdot \mathbf{\sigma }%
_{2})\mathbf{k}^{2}\right] \delta (\mathbf{r}_{1}-\mathbf{r}_{2})  \notag \\
&&+U\left[ (\mathbf{\sigma }_{1}\cdot \mathbf{k}^{\prime })\delta (\mathbf{r}%
_{1}-\mathbf{r}_{2})(\mathbf{\sigma }_{2}\cdot \mathbf{k})\right]   \notag \\
&&-\frac{U}{3}(\mathbf{\sigma }_{1}\cdot \mathbf{\sigma }_{2})\left[ \mathbf{%
k}^{\prime }\cdot \delta
(\mathbf{r}_{1}-\mathbf{r}_{2})\mathbf{k}\right] ,
\end{eqnarray}
where $\mathbf{k}=(\overrightarrow{\nabla
}_{1}-\overrightarrow{\nabla }_{2})/(2i)$ acts on the right and
$\mathbf{k}^{\prime }=-(\overleftarrow{\nabla }_{1}-\overleftarrow{\nabla }%
_{2})/(2i)$ acts on the left. $T$ and $U$ provide the intensity of
the tensor force in even and odd states of relative motion,
respectively. The spin-orbit potential including the tensor
contributions is given by \cite{SO1}
\begin{equation}
U_{\text{s.o.}}^{(q)}=\frac{W_{0}}{2r}\left( 2\frac{d\rho _{q}}{dr}+\frac{%
d\rho _{q^{\prime }}}{dr}\right) +\left( \alpha \frac{J_{q}}{r}+\beta \frac{%
J_{q^{\prime }}}{r}\right), \label{AA}
\end{equation}
where $q (q')$ denotes the like (unlike) particles and the $J_{q}$
is the proton or neutron spin-orbit density given as
\begin{equation}
J_{q}=\frac{1}{4\pi r^{3}}\underset{i}{\sum
}v_{i}^{2}(2j_{i}+1)\left[
j_{i}(j_{i}+1)-l_{i}(l_{i}+1)-\frac{3}{4}\right] R_{i}^{2}(r).
\end{equation}
where $v_{i}^{2}$ is the BCS occupation probability of each orbit.
The orbital with $j_{>}=l+1/2$ gives a positive contribution to
$J_{q}$ while the orbital with $j_{<}=l-1/2$ gives a negative
contribution to $J_{q}$. The strengths of these contributions
$\alpha$ and $\beta$ are expressed as $\alpha =\alpha _{C}+\alpha
_{T}$ and $\beta =\beta _{C}+\beta _{T}$. $\alpha _{C}$ and $\beta
_{C}$ are related to the central exchange part of the interaction
and can be written in terms of the usual Skyrme parameters
\begin{eqnarray}
\alpha _{C}
&=&\frac{1}{8}(t_{1}-t_{2})-\frac{1}{8}(t_{1}x_{1}+t_{2}x_{2}),
\nonumber\\
\beta _{C} &=&-\frac{1}{8}(t_{1}x_{1}+t_{2}x_{2})\text{.}
\end{eqnarray}
$\alpha _{T}$ and $\beta _{T}$ are related to the tensor part in
the following way
\begin{equation}
\alpha _{T}=\frac{5}{12}U,\text{\ \ \ }\beta _{T}=\frac{5}{24}(T+U).
\end{equation}
The central exchange and tensor contributions modify the energy
density $H$ by
\begin{equation}
\Delta H=\frac{1}{2}\alpha \left( J_{n}^{2}+J_{p}^{2}\right) +\beta
J_{n}J_{p}.
\end{equation}
One set of the parameters employed in the present work is the same
as that in Ref. \cite{SO1}: $\alpha _{T}=-170$ MeV fm$^{5}$ and
$\beta _{T}=100$ MeV fm$^{5}$ added perturbatively to the Skyrme
force SLy5 \cite{SLy5} (marked by SLy5+TF), which can fairly well
explain the isospin-dependence of the experimental energy
differences between the single-proton states outside the $Z=50$ core
for Sn isotopes, and the single-neutron states outside the $Z=82$
core for $Z=82$ isotones both quantitatively and qualitatively.
Recently, Zou \emph{et al}. analyzed the evolution of the spin-orbit
splittings in the Ca isotopes and $N = 28$ isotones with this
SLy5+TF and they found that adding the tensor contribution can
qualitatively explain in most cases the empirical trends
\cite{single2}.

\begin{figure}[htbp]
\begin{center}
\includegraphics[width=0.5\textwidth,clip]{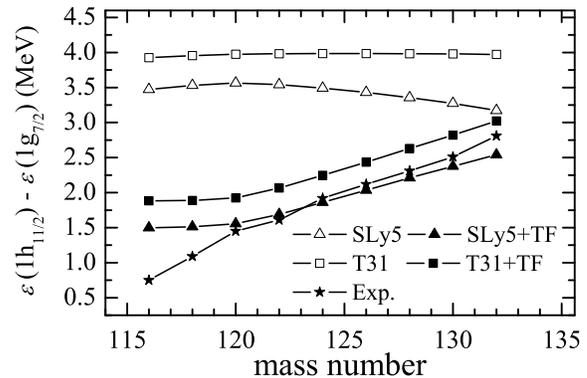}
\caption{Energy differences between the $1h_{11/2}$ and $1g_{7/2}$
single-proton states along the Sn isotopes. The experimental data
are taken from Ref. \cite{XXXX}. For the proton-rich isotopes, one
of or both of these two levels lie on the positive energy states
(unbound), so we do not present them here.}
\end{center}
\end{figure}

\section{Effect of the tensor force on the pseudospin energy splitting}\label{intro}\noindent
\begin{figure}[htbp]
\begin{center}
\includegraphics[width=0.5\textwidth,clip]{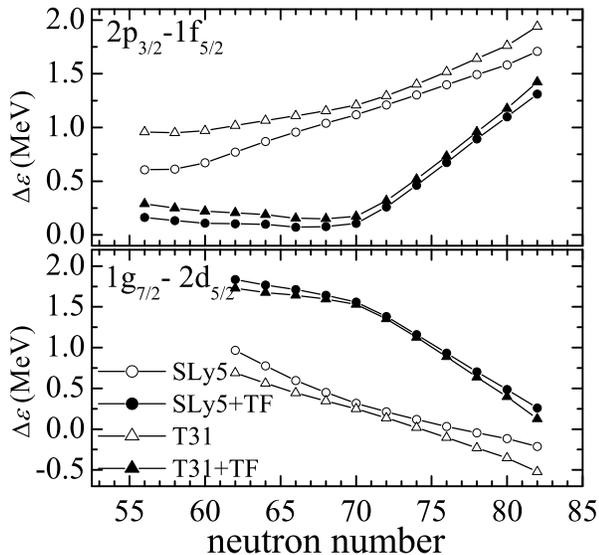}
\caption{Proton pseudospin orbit splitting $ \Delta \varepsilon$
measured by $\varepsilon (2p_{3/2})-\varepsilon (1f_{5/2})$ for
$1\widetilde{d}$ and by $\varepsilon (1g_{7/2})-\varepsilon
(2d_{5/2})$ for $1\widetilde{f}$ in the Sn isotopes.}
\end{center}
\end{figure}

\begin{figure}[htbp]
\begin{center}
\includegraphics[width=0.5\textwidth,clip]{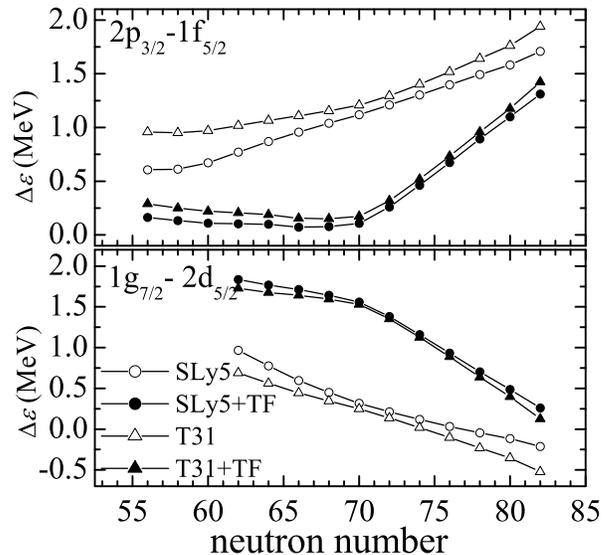}
\caption{Same as Fig. 1, but for neutrons.}
\end{center}
\end{figure}

Apart from affecting the spin symmetry, the tensor force influences
the pseudospin symmetry. The shift of single-particle levels might
have different origins and mechanisms. In the present study, we
focus on the effect of the tensor force. The calculations are
performed in a coordinate space using a box size of 20 fm and a mesh
size 0.1 fm. The pairing correlation is accounted in the BCS
formalism with an energy gap determined by the observed odd-even
mass differences for an open shell and being zero for a closed
shell. In addition to the SLy5+TF, the parameter set T31 including
the tensor force (T31+TF) is employed in the present study, which
also explains the experimental energy differences between the
$1h_{11/2}$ and $1g_{7/2}$ single-proton states along the Sn
isotopes, while the interaction T31 without the tensor component
fails to reproduce the trend of the experimental data, as shown in
Fig. 1. Note that the calculations with the SLy5+TF were firstly
performed in Ref. \cite{SO1} using other pairing formalisms, yet
their results coincide with our calculated ones, which is therefore
consistent with the conclusion that the pairing effect is negligible
for the spin-orbit splitting suggested in Ref. \cite{ZW00}. We
display the proton pseudospin energy splitting of $1\widetilde{d}$
$(2p_{3/2}, 1f_{5/2})$ and $1\widetilde{f}$ $(1g_{7/2}, 2d_{5/2})$
of the Sn isotopes in Fig. 2. The results with the parameter sets
SLy5 (SLy5+TF) and T31 (T31+TF) show the same trend. The proton spin
current does not contribute to isospin dependence of the spin-orbit
potential for an isotope chain, however, the neutron spin current is
uniquely responsible for the isospin dependence and affected greatly
by the sign of the parameter $\beta _{T}$. From $N=56$ to $N=64$,
$J_{n}$ is reduced as the $1g_{7/2}$ neutron orbit is gradually
filled. Because of a positive value of $\beta _{T}$, the absolute
values of the proton spin-orbit splittings are enlarged. Therefore,
the energy level of the proton orbit $1f_{5/2}$ is pushed up while
that of the $2p_{3/2}$ is pulled down, leading to the reduction of
the pseudospin orbit splitting of $1\widetilde{d}$. When the neutron
number $N$ changes from 66 to 70, the filling of the $2d_{3/2}$
neutron orbit also reduces the pseudospin orbit splitting of
$1\widetilde{d}$. The pseudospin energy splitting of
$1\widetilde{f}$ should increase in this region, nevertheless, the
lower panel of Fig. 2 shows an opposite trend. The reason for this
result is that the centroid of the $1g$ levels shifts down more
rapidly than that of the $2d$ levels as the neutron number
increases, which submerges the weak isospin-dependence of the tensor
effect in this region. Moreover, from $N=72$ to $N=82$, the
$1h_{11/2}$ orbit is gradually filled, which reduces the spin-orbit
splitting, so that the $1f_{5/2}$ and $1g_{7/2}$ orbits shift
downwards, but the $2p_{3/2}$ and $2d_{5/2}$ shift upwards.
Accordingly, the pseudospin energy splitting of $1\widetilde{d}$ is
enhanced, however, that of the $1\widetilde{f}$ is reduced as the
neutron number increases.

Although the proton spin current does not affect isospin dependence
of the spin-orbit potential for a given isotope chain, being
spin-unsaturated for protons, the filling of the proton orbits
affects nucleonic spin-orbit splitting as compared to that without
the tensor force. The negative value of $\alpha _{T}$ and positive
value of $J_{p}$ enhance the absolute value of the spin-orbit
potential and hence increase the spin-orbit splitting. From $N=56$
to $N=68$, the tensor effects resulting from the proton spin current
($\alpha _{T}$ term) and neutron spin current ($\beta _{T}$ term)
affect the spin-orbit splitting in the same way. Consequently, the
pseudospin energy splitting of $1\widetilde{d}$ is reduced and
$1\widetilde{f}$ is enlarged evidently compared to those without the
tensor force. From $N=70$ to $N=82$, the $\alpha _{T}$ and $\beta
_{T}$ terms vary in an opposite way, and the contribution of the
$\alpha _{T}$ term is cancelled out gradually. Accordingly, the
difference of the pseudospin energy splitting between that with and
without the tensor force is reduced gradually. On the whole,
compared to that without the tensor force, the pseudospin symmetry
is recovered to a large extent for the pseudospin doublet of
$1\widetilde{d}$ but is broken for $1\widetilde{f}$. The tensor
interaction thus brings a distinct change in pseudospin symmetry.
For neutrons, the role of the tensor force is more complicated due
to the variation of neutron number, but the obvious alternations
caused by the tensor component are distinguished, as shown in Fig.
3. The tensor force leads to the energy inversion of some pseudospin
doublets, such as $(2p_{3/2}, 1f_{5/2})$ when $N$ varies from $N=56$
to 60 as can be seen in Fig. 3.

\begin{figure*}[htbp]
\begin{center}
\includegraphics[width=1.0\textwidth,clip]{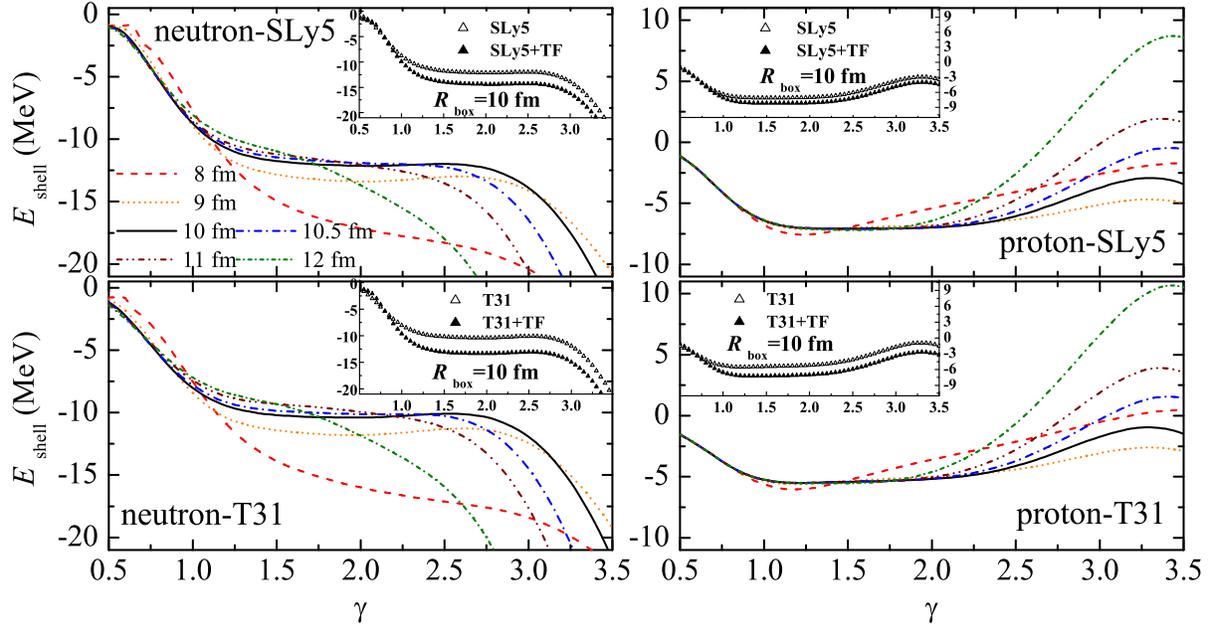}
\caption{Neutron and proton shell correction energies as a
function of the smoothing range $\gamma $ for $^{132}$Sn with
different box sizes. The insets display the shell correction
energies with and without the tensor component with the box size
$R_{\text{box}}=10$ fm and their coordinate axes are the same as
the figures. }
\end{center}
\end{figure*}

For the pseudospin doublets except (n$\widetilde{p}_{1/2}$,
n$\widetilde{p}_{3/2}$), such as (1$\widetilde{g}_{7/2}$,
1$\widetilde{g}_{9/2}$) and (1$\widetilde{h}_{9/2}$,
1$\widetilde{h}_{11/2}$) in the spin-unsaturated heavy nucleus, they
are all composed of a $j_{>}=l+1/2$ orbit and a $j_{<}=l-1/2$ orbit.
Accordingly, their pseudospin symmetry could be affected
considerably by the tensor effect because the two orbits are shifted
by the tensor component in an opposite direction. For the pseudospin
doublets (n$\widetilde{p}_{1/2}$, n$\widetilde{p}_{3/2}$), only the
n$\widetilde{p}_{3/2}$ orbit is shifted due to the tensor force (the
n$\widetilde{p}_{1/2}$ orbit is not affected due to its zero orbit
angular momentum), and hence the tensor force also affects these
pseudospin splittings. Our aim is to illustrate the significant
influence of the tensor force on the pseudospin symmetry. Because
the pseudospin symmetry is affected by the tensor force, a number of
phenomena \cite{DD,SD,MR1,MR2,band1,band2,band3,LWH} being related
to this symmetry can be accordingly affected. In addition, the
rotation bands of the deformed nuclei based on the single-particle
levels should be influenced by the tensor force, which needs
investigation but it is not going to be easy because of the breaking
of the time reversal symmetry.

\section{Effect of the tensor force on the shell correction}\label{intro}\noindent
Once the single-particle spectra are obtained, the shell correction
energies can be calculated. Calculating such a correction requires
the subtraction of the \textquotedblleft smoothing
varying\textquotedblright part of the sum of single-particle
energies
\begin{equation}
E_{\text{shell}}=E-\widetilde{E}=\overset{N(Z)}{\underset{i=1}{\sum }}%
e_{i}n_{i}-\int_{-\infty }^{\widetilde{\lambda
}}e\widetilde{g}(e)de,
\end{equation}
where $N(Z)$ is the neutron (proton) number, and $e_{i}$ are the
single-particle energies. The neutrons and protons are treated quite
separately. The smoothed Fermi level $\widetilde{\lambda }$ is
obtained from the equation $N(Z)=\int_{-\infty }^{\widetilde{\lambda
}}\widetilde{g}(e)de$. Taking into account the $n_{i}$-fold
degeneracies of the single particle orbits, the smoothed level
density $\widetilde{g}(e)$ takes the form of
\begin{equation}
\widetilde{g}(e)=\frac{1}{\gamma }\overset{\infty }{\underset{i=1}{\sum }}%
n_{i}f\left( \frac{e_{i}-e}{\gamma }\right) ,
\end{equation}
where $\gamma $ is the smoothing range in units of $\hbar \omega
_{0}=41A^{-1/3}$\ MeV. The folding function is usually taken as
$f(x)=\frac{1}{\sqrt{\pi }}e^{-x^{2}}L_{M}^{1/2}(x^{2})$, where
$L_{M}^{1/2}(x^{2})$ is an associated Laguerre polynomial. Here
the energy cutoff of the single-particle spectra is selected to be
40 MeV.

\begin{figure}[htbp]
\begin{center}
\includegraphics[width=0.5\textwidth,clip]{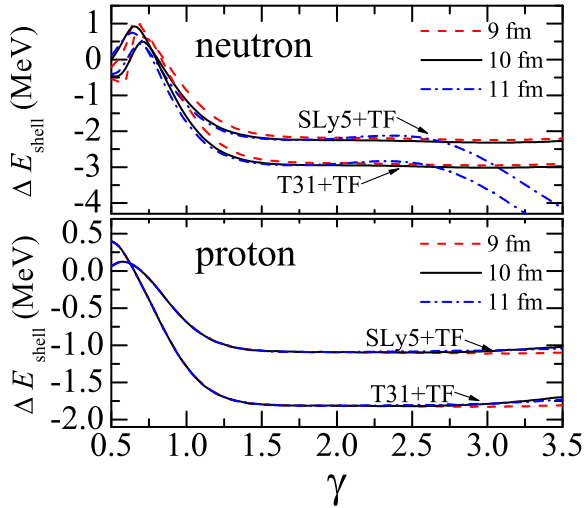}
\caption{Modification of neutron and proton shell correction
energies due to the tensor force as a function of the smoothing
range $\gamma $ for $^{132}$Sn.}
\end{center}
\end{figure}

It remains to determine the order of the associated Laguerre
polynomial $M$ and the smoothing range $\gamma $. Ideally, the
calculated $E_{\text{shell}}$ should be not dependent on the
specific values in a broad range of reasonable values, namely, the
so-called plateau condition
\begin{equation}
\frac{\partial E_{\text{shell}}}{\partial \gamma }=0,\text{ \ \ }\frac{%
\partial E_{\text{shell}}}{\partial M}=0.
\end{equation}
In our calculations, it is found that $M=3$ is an optimal value just
like that in Ref. \cite{NYF}, in which the associated Laguerre
polynomial takes the form of
$L_{3}^{1/2}(x^{2})=-x^{6}/6+7x^{4}/4-35x^{2}/8+35/16$. Fig. 4
presents the neutron and proton shell correction energies as a
function of the smoothing range $\gamma $ taking the nucleus
$^{132}$Sn as an example, where the Skyrme-Hartree-Fock equations
are solved with box sizes of $R_{\text{box}}=8\sim 12$ fm with the
parameter sets SLy5 and T31. One can find that the plateau condition
is well fulfilled within the range $1.5\leqslant \gamma \leqslant
2.7$ for neutrons and $1.2\leqslant \gamma \leqslant 2.0$ for
protons in the case of $R_{\text{box}}=10$ fm. The plateau vanishes
gradually as $R_{\text{box}}>10$ fm. And $R_{\text{box}}=10$ fm is
large enough in our calculations to achieve the convergence for the
total binding energy with an accuracy of $0.03\%$. When
$R_{\text{box}}=10$ fm, the calculated shell correction energies for
$^{132}$Sn with the parameter sets SLy5 and T31 in the cases with
and without the tensor force, are presented in the insets of Fig. 4.
This shell correction shows a parameter dependence. Without the
tensor force, the shell correction energies for neutron (proton) are
about -12.0 MeV (-7.0 MeV) and -10.3 MeV (-5.5 MeV) by adopting the
parameter sets SLy5 and T31, respectively. The tensor force shifts
the single-particle levels, thereby the single-particle level
density undergoes a corresponding change, so that the shell
correction energy is modified. It can be found that the shell
correction energies for both neutrons and protons are reduced by the
tensor force, the reason for which is partly because the tensor
interaction makes the energy level density near the Fermi surface
become lower. In order to show a clearer change in the shell
correction energy caused by the tensor force, we plot in Fig. 5 the
net effect of the tensor force on the shell correction energy
$\Delta E$, namely, the difference between the shell correction
energies with and without the tensor force. As can be seen, the
$\Delta E$ presents a more broad range of $\gamma$ from 1.5 to 3.5
to fulfill the \textquotedblleft plateau condition\textquotedblright
with $R_{\text{box}}=10$ fm. With $R_{\text{box}}=11$ fm, though the
neutron and proton shell correction energies do not show the
\textquotedblleft plateau \textquotedblright as presented in Fig. 4,
the net contribution of the tensor force to the shell correction
energies shows the \textquotedblleft plateau \textquotedblright,
especially for protons and nearly provides the same result as that
with $R_{\text{box}}=10$ fm. This indicates the reliability of our
calculations to a large extent. The error caused by the box size can
be cancelled out to a large extent when one computes the
contributions from the tensor force. One can find that due to the
tensor force, the neutron shell correction energy is lowered by
about 2.3 MeV and 3.0 MeV, and the proton shell correction energy is
lowered by about 1.1 MeV and 1.8 MeV with the parameter sets SLy5+TF
and T31+TF, respectively. In other words, the effect of the tensor
force on the shell correction is considerable. By the way, the shell
correction energy at the saddle point turns out to be too important
to be neglected \cite{ZW}, so the tensor force may play a
non-negligible role in fission processes.

\begin{figure}[htbp]
\begin{center}
\includegraphics[width=0.5\textwidth,clip]{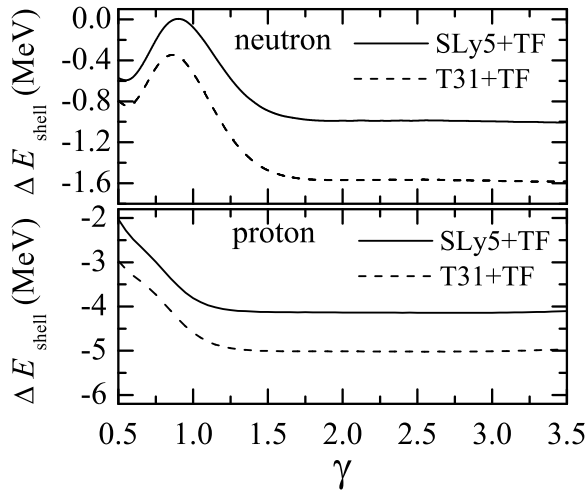}
\caption{Modification of neutron and proton shell correction
energies due to the tensor force as a function of the smoothing
range $\gamma $ for $^{298}$114.}
\end{center}
\end{figure}
\begin{figure}[htbp]
\begin{center}
\includegraphics[width=0.5\textwidth,clip]{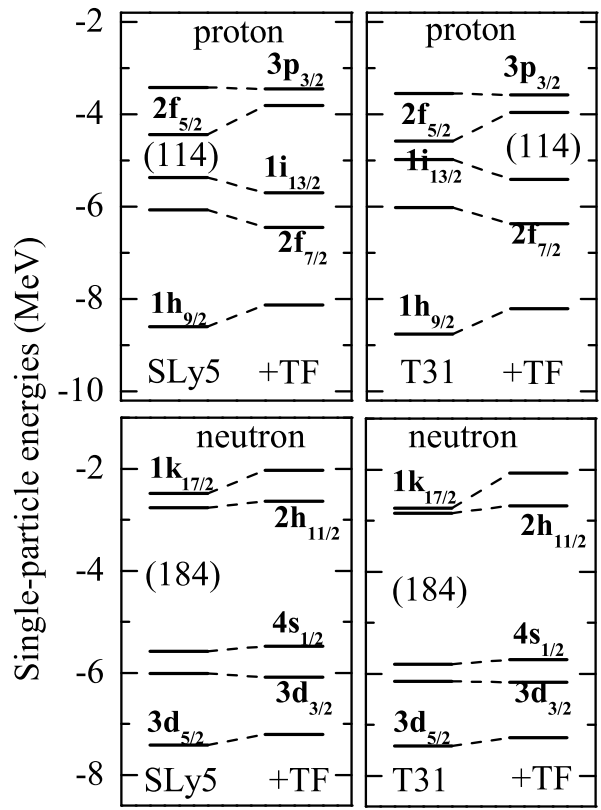}
\caption{Single-particle spectra of $^{298}$114 near the Fermi
surface. The calculations are performed with (SLy5+TF and T31+TF)
and without (SLy5 and T31) the tensor interaction in the
spin-orbit potential.}
\end{center}
\end{figure}

Now we turn to the shell correction for superheavy nuclei. The
existence of the superheavy island was predicted in 1960s. However,
there is no consensus among theorists with regard to what should be
the next doubly magic nucleus beyond $^{208}$Pb. Nearly all of
modern calculations predict the existence of a closed neutron shell
at $N=184$. However, they differ in predicting the position of the
closed proton shell. Modern calculations predict that they appear at
$Z = 114$, 120, 124, or 126 \cite{HF,PP1,AB,SHF2,RMF,SHF3,RMF0}
depending on the models employed. We investigate the tensor effect
on the shell correction of superheavy nuclei taking $^{298}$114 as
an example. The influence of the tensor force on the neutron and
proton shell correction energies ($R_{\text{box}}=12.5$ fm) is
illustrated in Fig. 6. With the parameter sets used above, we find
that the shell correction energy for protons is altered by $4\sim 5$
MeV due to the inclusion of the tensor component, and is much larger
than the $1.0\sim 1.6$ MeV for neutrons. This implies that
$^{298}$114 becomes more stable due to the presence of the tensor
force with the present parameter sets since the shell correction
energy can reflect the stability of nuclei. We plot in Fig. 7 the
single-particle levels near the Fermi surface to show the effect of
the tensor force ($R_{\text{box}}=20$ fm). As can be seen, the shell
gap at $Z=114$ is obviously enlarged due to the tensor force, and
the shell gap at $N=184$ is changed slightly, which is consistent
with the above discussion about the shell correction energy and is
agreement with the results in Ref. \cite{SHN}. The difference
between the $4s_{1/2}$ and $2h_{11/2}$ neutron orbits determines the
size of the gap at $N=184$, yet the tensor force only directly
affects the $2h_{11/2}$ orbit. In consideration of the fact that the
$\alpha_{T}$ and $\beta_{T}$ terms in Eq. (\ref{AA}) take effect in
an opposite way, their contributions to the spin-orbit splitting are
cancelled out to a certain extent, hence the predicted neutron shell
structures is modified only slightly by inclusion of the tensor
component. However, for $Z=114$, the two proton orbits $2f_{5/2}$
and $1i_{13/2}$ shift in an opposite direction due to the tensor
component, thus the gap is changed obviously. The discussions here
aim at displaying the effect of the tensor force on the structure
and stability of the superheavy nucleus. The theoretical results at
least imply that the tensor force should play an important role in
the shell structure of the superheavy nucleus. We would like to say
that many-body approaches with various parameter sets tend to
predict different locations of the shell gaps, which perhaps results
from the uncompleted knowledge about the tensor force to a large
extent.

\section{Summary}\label{intro}\noindent
A detailed investigation for the effect of the tensor force on the
pseudospin energy splitting have been carried out taking Sn isotope
chain as an example in the framework of the Skyrme-Hartree-Fock
approach with the parameter sets SLy5+TF and T31+TF combined with
BCS method, where the sign of $\alpha _{T}$ and $\beta _{T}$ plays a
crucial role. The contributions of the tensor force to the
pseudospin symmetry are significant since all pseudospin doublets
except (n$\widetilde{p}_{1/2}$, n$\widetilde{p}_{3/2}$) are composed
of a $j_{>}$ orbit and a $j_{<}$ orbit which are affected in an
opposite way by the tensor force. For (n$\widetilde{p}_{1/2}$,
n$\widetilde{p}_{3/2}$), only n$\widetilde{p}_{3/2}$ orbit is
shifted due to the tensor force, and hence the tensor force also
affects the pseudospin splitting. Thus a lot of phenomena being
related to the pseudospin symmetry should be accordingly affected.
In addition, the effect of the tensor force on the shell correction
energy is calculated taking a magic nucleus $^{132}$Sn and a
superheavy nucleus $^{298}114$ as examples. It was shown that the
influence of the tensor force on the shell correction energy is
considerable. For $^{298}114$, the absolute values of calculated
proton shell correction energy is enlarged due to the inclusion of
the tensor component, which is consistent with the result that the
shell gap at $Z=114$ is obviously enhanced with the present
parameter sets. The modification of the shell correction energies by
the tensor component are important for superheavy nuclei, which are
related to their stability.  All the conclusions here actually
originate from the shifts of the single-particle levels (i.e.
modified spin-orbit splittings) on account of the presence of the
tensor force.

\section{ACKNOWLEDGMENTS}
J. M. Dong would like to thank Profs. W. H. Long and S. G. Zhou for
their helpful discussions. This work was supported by the National
Natural Science Foundation of China (with Grant Nos.
10875151,10575119,10975190,10947109), the Major State Basic Research
Developing Program of China under Grant Nos. 2007CB815003 and
2007CB815004, the Knowledge Innovation Project (KJCX2-EW-N01) of
Chinese Academy of Sciences, CAS/SAFEA International Partnership
Program for Creative Research Teams (CXTD-J2005-1), the Fundamental
Research Funds for the Central University under Grant No.
Lzujbky-2010-160 and the Funds for Creative Research Groups of China
under Grant No. 11021504.

\end{document}